\documentclass[useAMS,usenatbib,usegraphicx]{mn2e} 
\usepackage{aas_macros}

\title[Five nights of intensive $R$- and $V$-band photometry of
  QSO~0957+561A,B]
  {Five nights of intensive \textit{R}- and \textit{V}-band photometry
  of QSO~0957+561A,B}  
\author[J.~E.~Ovaldsen et al.]
{J.~E.~Ovaldsen,$^1$\thanks{E-mail: j.e.ovaldsen@astro.uio.no}
  J.~Teuber,$^2$  R.~Stabell$^1$ and A.~K.~D.~Evans$^1$\\
$^{1}$Institute of Theoretical Astrophysics, University of Oslo,
  P.~O.~Box 1029, Blindern, N-0315 Oslo, Norway \\
$^{2}$Centre for Advanced Signal Processing, Copenhagen, Denmark}

\begin{document}

\date{Accepted 2003 July 8. Received 2003 May 13}

\pagerange{\pageref{firstpage}--\pageref{lastpage}} \pubyear{2003}
\maketitle
\label{firstpage}

\begin{abstract}
  We present $R$- and $V$-band photometry of the gravitational lens
  system QSO 0957+561 from five nights (one in 2000 January and four
  in 2001 March, corresponding to the approximate time-delay for the
  system) of uninterrupted monitoring at the Nordic Optical Telescope.
  In the photometry scheme we have stressed careful magnitude
  calibration as well as corrections for the lens galaxy contamination
  and the crosstalk between the twin (A and B) quasar images. The
  resulting, very densely sampled light curves are quite stable, in
  conflict with earlier claims derived from the same data material.
  We estimate high-precision timelag-corrected B/A flux ratios in both
  colour bands, as well as $V-R$ colour indices for A and B, and
  discuss the short time-scale variability of the system.
\end{abstract}



\section{Introduction}
More than 20 years of monitoring of the `classical' double quasar QSO
0957+561A,B \citep{Walsh79} has shown that this interesting object
exhibits variability at several levels and time-scales.  This
variability has been extremely useful for such diverse applications
as, e.g., estimating the Hubble constant via the `Refsdal time delay
method', constraining the continuum-emitting region of the quasar, and
probing the mass distribution in the lens galaxy and its surrounding
cluster.  Despite its relative photometric simplicity, QSO 0957+561
($z_\mathrm{QSO}=1.41$, $z_\mathrm{lens}=0.36$, $\theta_\mathrm{AB}
\approx 6.2$ arcsec) has nevertheless proved more unwieldy than
initially hoped.  In this context we note that even the best sampled
light curves may exhibit artificial fluctuations or seeming
correlations, and the task of distinguishing between physical events
(whether intrinsic or microlensing-induced) and spurious ones is not
necessarily straightforward.

In this paper we employ our newly developed photometry scheme
\citep{OTSS03} and present two-colour photometry for QSO 0957+561A,B
based upon about 1200 CCD images from five observing nights, four of
which are consecutive.  This densely sampled, high-quality data set
allows us to, e.g., constrain the short time-scale variability and the
noise level to be expected in past and future investigations on the
same theme. Part of this study constitutes a re-reduction of existing
data, and our results are at variance with previously published work
\citep[][ hereafter C02 and C03]{Colley02,Colley03}; for instance, we
find no support for the hourly variability invoked by the above
authors to facilitate a time delay determination with accuracy of a
fraction of a day (417.09$\pm$0.07 d).

The data material is described in Sect.~\ref{S:data}, while a short
outline of the photometric procedure is given in Sect.~\ref{S:phot}.
Photometry for comparison stars as well as for the A and B quasar
components is presented and commented in Sect.~\ref{S:results}.
Sect.~\ref{S:D&S} concludes with a discussion and a summary.

\section{Data set} \label{S:data}
Our original data set consists of 976 $R$-band and 229 $V$-band CCD
frames obtained with the 2.56-m Nordic Optical Telescope (NOT) on La
Palma, Canary Islands, Spain, using the ALFOSC
instrument.\footnote{ALFOSC is owned by the Instituto de Astrofisica
  de Andalucia (IAA) and operated at the NOT under agreement between
  IAA and the NBIfAFG of the Astronomical Observatory of Copenhagen.}
The observations were made on 2000 January 25--26 and during the four
nights 2001 March 14--18, and were part of the `QuOC Around the Clock'
2000/2001 campaigns, designed to compare observing runs separated by
the assumed time delay for QSO 0957+561 of $\sim$416 days; see C02 and
C03, where a reduction of the $R$-band data from the same set is
discussed. (Note that investigations in recent years seem to favour a
lag around 425 days, see \citealp{OTSS03} and references therein.) The
detector is a 2k$\times$2k CCD with a gain of 1~e$^{-}$/ADU and
readout noise of approx.\ 6~e$^{-}$/pixel.  The nominal pixel size is
0.188 arcsec.

All images were pre-processed with bias subtraction and flat-fielding,
exclusively by means of sky flats.  The averaged flats typically
represented a photon number 100--500 times the background level (which
was between 100 and 600 photons) in the science frames.  According to
the online specification of the CCD chip, it contains 8 bad columns,
10 traps, and many dark `speckles'. Deplorably, we had to discard 107
frames ($97\,R+10\,V$) because one or both of the quasar components
fell too close to a defective column.

The seeing varied from 0.6 to 2 arcsec (median value $\approx 0.9$).
The point sources on the frames were usually slightly elliptical, with
a median axis ratio of 1.1.  However, some PSFs were highly
elliptical, especially when the object was near the horizon.

\section{Photometry scheme} \label{S:phot}
The photometry package used to reduce the CCD images was developed by
us (JT and JEO). The method is described in \citet{OTSS03}, where it
was applied to a five-year archival data set of QSO 0957+561; a full
treatment is given in \citet{Ovaldsen02}.  Features include automatic
source detection and localisation, field star aperture and PSF
photometry, magnitude calibration, and aperture and PSF photometry of
the twin quasar images (after lens galaxy subtraction and crosstalk
correction).

\subsection{Field star photometry and calibration} 
To obtain a reliable magnitude calibration, we believe it is
imperative to include as many comparison stars as possible \emph{and}
to check their internal brightness consistency.  We apply aperture
photometry, using a 3 arcsec radius, to 7 comparison stars in the
field (F, G, H, E, D, X, and R, see fig.~1 in \citealp{OTSS03}).  The
measured total intensities of the field stars are compared to the
reference values (see below) to estimate the frame conversion factor.
If more than one outlier is present, or if the conversion factor is
statistically ambiguous, the frame is discarded.  The obvious
rationale for this policy is that if consistent photometry for the
reference stars is impossible, it will be even more so for the quasar
components.

Reference $R$-magnitudes for the comparison stars were kindly supplied
by R.\ Schild (private communication).  Lacking a proper set of
reference $V$-magnitudes, we used the $V-R$ colour terms of
\citet{Serra99} for the stars H and D to convert our $R$-magnitudes
for these two stars into $V$. Finally, the overall $R$- and
$V$-magnitude scales were adjusted to a best fit with the reference
values.

It should be noted that the above procedure implies a large difference
between our reference $R$-magnitudes and those of \citeauthor{Serra99}
We would like to stress, however, that this is not unusual.  The
reference magnitudes encountered in the literature (e.g.\ 
\citealp{Keel82}, \citealp{Vander89}, and \citeauthor{Serra99}), may
differ by as much as 0.4 mag for a single star!

\subsection{Photometry of the quasar images} \label{S:qso_phot}
The distinctive feature of QSO 0957+561 is the seeing-dependent
contamination by the lensing galaxy, denoted G1, particularly of the
closely juxtaposed B image. To correct for this influence, we
construct and subtract from each frame a galaxy model.  Using a
composite $R$-band brightness profile derived by \citet{Bern97} and
the appropriate ellipse parameters, a seeing-free G1 model is
synthesised and convolved with the PSF of the actual frame (see
\citealp{Ovaldsen02} for details, especially as regards the magnitude
calibration). The position and orientation is determined using the
astrometry of \citeauthor{Bern97} and the calculated centroids of the
quasar images.  Note that the brightness level is adjusted by an
amount estimated by a three-component (A, B, G1) simultaneous fitting
in the entire data set. Without better options (see, however,
\citealp{Teuber03}), we use the same galaxy profile in the $V$-band,
decreasing the brightness level accordingly.

After the galaxy subtraction, the PSF is used to correct for
cross-contamination between the quasar components. Refraining from
performing a PSF flux estimation as such, we have chosen the hybrid
method consisting of a `PSF cleaning' of one of the two quasar images
followed by an aperture flux measurement of the other, and vice versa.
If necessary, the procedure may be performed iteratively.

Separate corrections for galaxy contamination and aperture crosstalk
are vital for achieving high-precision photometry, especially when there
are large variations not only in seeing but also PSF ellipticity.

\section{Photometric results} \label{S:results}
\subsection{Field stars} \label{S:star_calib}
Our calibration procedure accepted 817 (out of 879) $R$-band frames
and 206 (out of 219) $V$-band frames.  Photometry results for the
comparison stars are summarised in Tables~\ref{t:stat_R} and
\ref{t:stat_V} for the $R$- and $V$-band, respectively.

When presenting the quasar light curves (Sect.~\ref{S:quasar}), we
include the results for the fainter star, R, the brightness of which
is comparable to that of the quasar components.

\subsection{Galaxy contribution and aperture crosstalk}
From the subtraction procedure described above, we found that,
depending on seeing, the galaxy light in the $R$-band amounted to
18.24--18.26 and 20.45--20.25 mag in the B and A apertures,
respectively. In the $V$-band, the corresponding values were
19.34--19.38 mag (B) and 21.50--21.30 mag (A). So, as seeing
deteriorates, galaxy light spills out of the B aperture and into the A
aperture. Crosstalk between the twin quasar images was less than 1~per
cent for seeing values below 2 arcsec. Note that both corrections are
functions also of PSF ellipticity, that is, axis ratio and position
angle.

\begin{table}
\caption{Aperture photometry of the comparison stars in the
$R$-band. For each star the mean magnitude and standard deviation are
tabulated, along with the mean of the individual formal errors, and
reference magnitude. See text for explanation of the calibration.}
\begin{tabular}{@{}ccccc@{}} \hline
Star & Mean $R$-mag & $\sigma_\mathrm{mag} $ & Mean formal error & 
Ref.~mag \\ \hline 
 F  &  13.759  &  0.0016  &  0.0013  &  13.758 \\
 G  &  13.731  &  0.0013  &  0.0013  &  13.731 \\
 H  &  13.973  &  0.0016  &  0.0015  &  13.972 \\
 E  &  14.778  &  0.0020  &  0.0020  &  14.778 \\
 D  &  14.513  &  0.0023  &  0.0018  &  14.513 \\
 X  &  13.425  &  0.0014  &  0.0011  &  13.425 \\
 R  &  16.328  &  0.0046  &  0.0045  &  16.329 \\ \hline
\end{tabular}
\label{t:stat_R}
\end{table}
\begin{table}
\caption{Same as Table~\ref{t:stat_R}, but in the $V$-band.}
\begin{tabular}{@{}ccccc@{}} \hline
Star & Mean $V$-mag & $\sigma_\mathrm{mag} $ & Mean formal error & 
Ref.~mag \\ \hline 
 F  &  14.093  &  0.0015  &  0.0014  &  14.093 \\
 G  &  14.054  &  0.0013  &  0.0013  &  14.054 \\
 H  &  14.497  &  0.0016  &  0.0016  &  14.497 \\
 E  &  15.247  &  0.0021  &  0.0021  &  15.247 \\
 D  &  14.948  &  0.0021  &  0.0019  &  14.948 \\
 X  &  13.770  &  0.0014  &  0.0011  &  13.770 \\
 R  &  17.088  &  0.0061  &  0.0060  &  17.086 \\ \hline
\end{tabular}
\label{t:stat_V}
\end{table}

\subsection{The A and B quasar light curves} \label{S:quasar}
\begin{figure}
\includegraphics[width=84mm]{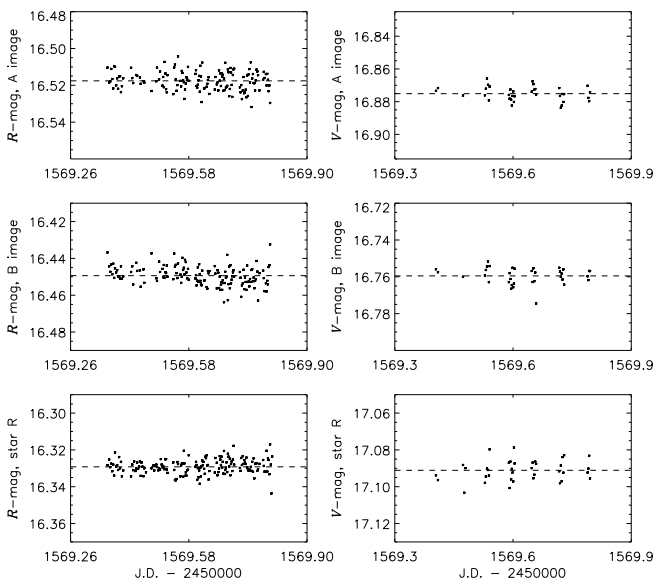}
\caption{Results from photometry of QSO 0957+561A,B and star~R in the
  $R$-band (left panels) and $V$-band (right panels) on 2000 January
  25--26. There are 144/35 ($R$/$V$) unbinned data points for each
  quasar image. Dashed lines are the mean level.}
\label{f:AB_RV2000}
\end{figure}

\begin{figure*}
\includegraphics[width=17cm]{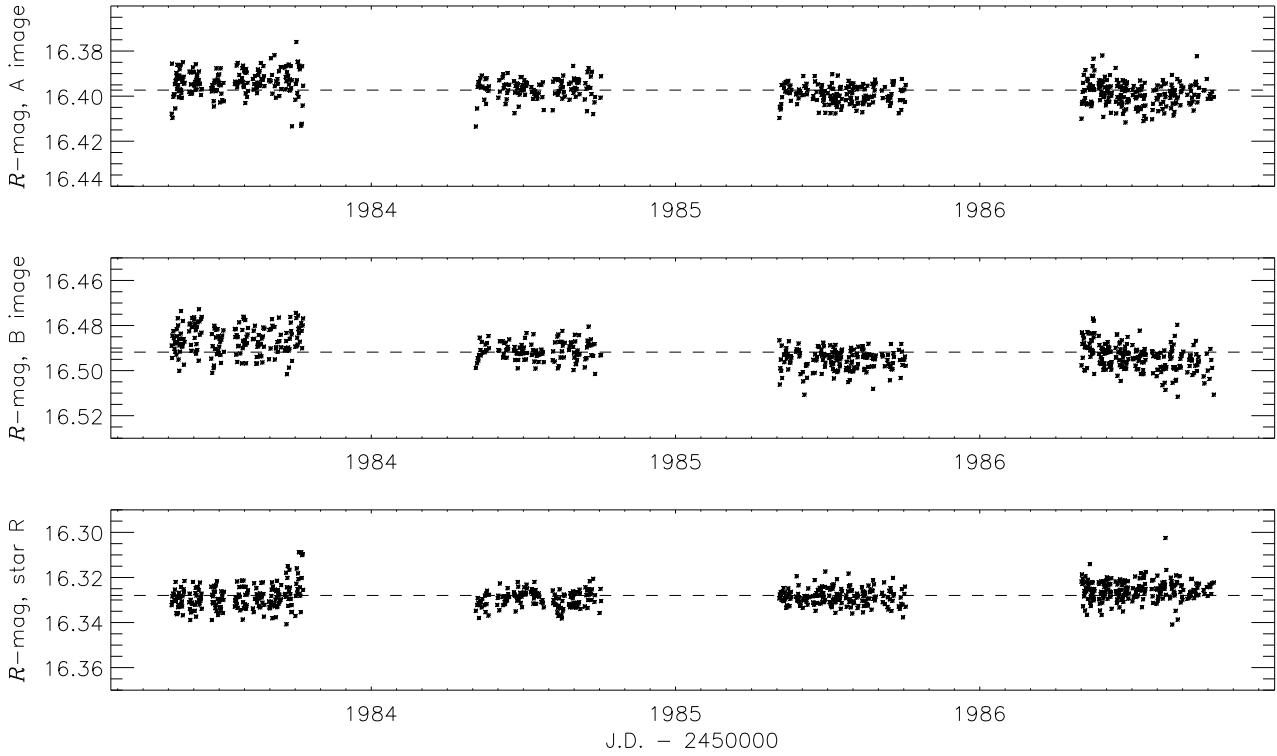}
\caption{Results from photometry of QSO 0957+561A,B and
  star~R in the $R$-band. Civil dates: 2001 March 14--18.  There are
  673 unbinned data points for each quasar image.  The mean of all
  data points is plotted as a horizontal dashed line, and is only
  meant as a guide to the eye.}
\label{f:AB_Rmags}
\end{figure*}

\begin{figure*}
\includegraphics[width=17cm]{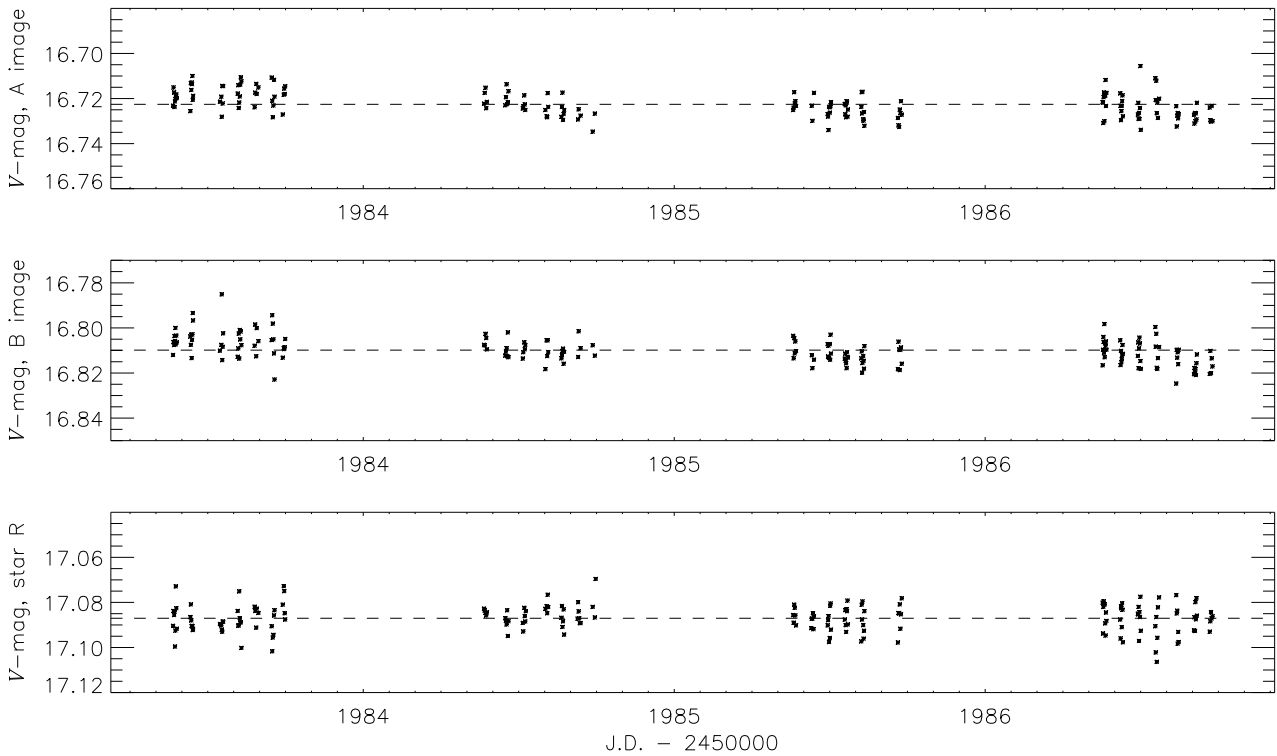}
\caption{Same as Fig.~\ref{f:AB_Rmags}, but in the
  $V$-band. Here there are 171 unbinned data points for each quasar image.}
\label{f:AB_Vmags}
\end{figure*}

Figs.~\ref{f:AB_RV2000} (2000 January), \ref{f:AB_Rmags} and
\ref{f:AB_Vmags} (2001 March) display light curves of the A and B
quasar images and the R star in the $R$- and $V$-bands.  No outliers
have been removed.  The nightly mean brightnesses of the same objects,
along with the corresponding standard deviations, are given in
Table~\ref{t:qso_stats}.

%
%
\begin{table*}
\centering
\begin{minipage}{170mm}
\caption{Mean $R$- and $V$-magnitudes for the A and B quasar components
  and the R star for the 5 nights (see Figs.\ \ref{f:AB_RV2000},
  \ref{f:AB_Rmags} and \ref{f:AB_Vmags}). Numbers in parentheses are
  the standard deviations in mmag.}
\begin{tabular}{crccccc} \hline
Band & Object & 2000 Jan.\ 25--26 \hspace{3mm} & 2001 March 14--15 &
    2001 March 15--16 & 2001 March 16--17 & 2001 March 17--18 \\ \hline
 & A & 16.518 (5.2) \hspace{3mm} & 16.394
   (5.7) & 16.397 (4.4) & 16.399 (4.0) & 16.399 (5.1) \\
$R$ & B & 16.449 (5.3) \hspace{3mm} & 16.486
   (6.2) & 16.491 (4.0) & 16.495 (4.4) & 16.494 (5.7) \\
 & Star R & 16.329 (4.6) \hspace{3mm} & 16.329
   (5.4) & 16.329 (3.5) & 16.329 (3.5) & 16.326 (4.3) \\
\hline
 & A & 16.875 (4.3) \hspace{3mm} & 16.718
   (4.7) & 16.723 (4.7) & 16.725 (5.2) & 16.724 (5.4) \\
$V$ & B & 16.760 (4.5) \hspace{3mm} & 16.806
   (6.4) & 16.810 (3.9) & 16.812 (4.3) & 16.812 (5.8) \\
 & Star R & 17.091 (5.6) \hspace{3mm} & 17.088
   (6.3) & 17.086 (4.8) & 17.087 (5.3) & 17.088 (6.8) \\
\hline
\end{tabular}
\label{t:qso_stats}
\end{minipage}
\end{table*}

Despite considerable hourly variations of such decisive observational
parameters as airmass, seeing, PSF ellipticity, sky level, and
S/N-ratio, the light curves exhibit no dramatic features and appear
quite constant during each night. The nightly standard deviations for
the quasar components range from 3.9 to 6.4 mmag, to be compared with
the mean of the individual formal errors for each night of 4.6 to 7.1
mmag -- in satisfactory agreement with the corresponding values for
the reference R star. The formal uncertainties are rather
conservative, as they include the errors from Poisson statistics,
galaxy subtraction and magnitude calibration (see \citealp{Ovaldsen02}
for details).  We omit plotting error bars in the figures to increase
clarity.

The most conspicuous features of the brightness data are the changes
in A and B over a time span of the order of one year.  It is found
that A brightens by 0.12 mag in $R$ and 0.15 mag in $V$, whereas B
becomes fainter by 0.043 mag in $R$ and 0.050 mag in $V$.

From Figs.~\ref{f:AB_Rmags} and \ref{f:AB_Vmags} and
Table~\ref{t:qso_stats}, one might also deduce a small brightness
decrease in A and B, in both colours, of 1--4 mmag/day. However, the
individual data sets for each of the four consecutive nights reproduce
little, if any, of this overall trend (see Sect.~\ref{S:D&S}).  For a
discussion of the quasar variability over several weeks, see the
recent paper by \citet{Ullan03}.

As the observing runs were separated by the approximate time delay of
416 days, we are now able to compare the brightness values of the A
quasar component on 2000 January with the corresponding time-shifted
ones of B (average of the 2001 March values).  We find $V-R$ colour
indices of 0.357$\pm$0.007 mag for A and 0.318$\pm$0.009 mag for B.
Converting the $m_\mathrm{B}-m_\mathrm{A}$ differences from
Table~\ref{t:qso_stats} into the physically significant B/A
magnification ratios, we find values of 1.025$\pm$0.011 in $R$ and
1.062$\pm$0.009 in $V$.  These results are in qualitative agreement
with \citet{Serra99}, but not quite consistent with the achromaticity
of the gravitational lens and the expected differential reddening of B
due to its larger optical depth (see also Sect.~\ref{S:D&S}).

\subsection{Comparison with previous reductions}
As regards the quasar variability in the two periods, our results --
typically constant light curves in both colours during each night --
do not support the many short time-scale fluctuations seen in fig.~1
of C03 (and C02). We therefore suspect that most of these fluctuations
may be spurious and a likely reflection of problems arising from
combining photometry from 12 different observatories.
Likewise, failure to reject images of poor quality -- whether due to
image flaws or, more subtly, to inconsistent reference star photometry
-- will inevitably introduce systematic errors which may easily
be misidentified as brightness fluctuations.

The A and B brightness levels for both observing runs, 2000 January
and 2001 March, also differ significantly from those of C02 and C03.
As an example, we find that during the observing nights in the
$R$-band 2001 March, A and B were on average 16.397 and 16.492 mag,
respectively, i.e.\ A brighter than B. C03 (fig.~1) finds
$\sim\!\!16.37$ (A) and $\sim\!\!16.30$ (B), i.e.\ B brighter than A.
The above authors perform a single correction, solely a function of
seeing, to account for both galaxy light contamination and aperture
crosstalk.  If, however, we \emph{omit} the galaxy subtraction and
only apply our crosstalk correction, we get average values for A and B in
perfect agreement with C03. (Note that our corrections not only
account for seeing effects on each particular frame, but also for PSF
ellipticity.)

\section{Discussion and summary} \label{S:D&S}
A careful reduction with our new photometry scheme has yielded quasar
light curves with little noise and hardly any outliers among the more
than 2000 A and B data points. The photometry program seems to cope
well with the large nightly variations in airmass, seeing, PSF
ellipticity, sky level, and S/N-ratio, and produces remarkably
constant light curves for each night. Our light curves do not exhibit
any hourly fluctuations, in contrast to what e.g.\ \citet{CS00}, C02
and C03 have found.  As a consequence, it seems impossible to derive
any firm time delay from the observational material discussed here,
let alone to obtain an estimate accurate to five significant digits,
as the authors of C03 have been able to.

The small decrease from one night to the next for both quasar images,
in addition to the typically constant behaviour during each night,
might be regarded as a tentative, short time-scale manifestation of
the much-debated `zero lag correlation' (between the theoretically
unrelated A and B light curves) present in the brightness record of
QSO 0957+561 \citep{Kundic95,Colley03,OTSS03}. In the last reference,
several conspicuous, simultaneous fluctuations were found in the
quasar light curves, with time-scales of up to several months and
amplitudes up to 0.2 mag. However, four nights of continuous
monitoring do admittedly not suffice to draw any firm conclusions
about this issue. It constitutes nevertheless a strong indication that
the correlation is a hitherto unexplained \emph{observational} effect.

If not due to intrinsic fluctuations or microlensing, the observed
small brightness variations -- and the possible zero lag correlation
as well -- might be ascribed to colour effects arising from, e.g.,
changing observing conditions in connection with the spectral mismatch
between quasar and comparison stars. However, the same remarks as in
Sect.~\ref{S:quasar} regarding the large nightly changes in
observational parameters coupled with the constant nightly light
curves render this explanation unlikely. It remains a puzzle, though,
that the night-to-night trend is not reflected in the nightly
observations.  Likewise, \citet{Ullan03} found unexplained day-to-day
variations that were very similar in both A and B. Because we have
both constant and densely sampled light curves during the course of
each night, we can exclude at least seeing, ellipticity, airmass and
sky level from the list of possible explanations.

Combining our 2000 January A data with the 2001 March B data, in both
colour bands, we have been able to estimate the optical continuum
magnification ratio of QSO 0957+561A,B as well as the colour indices
of the quasar images. The results stated in Sect.~\ref{S:quasar} seem
somewhat surprising. There is no obvious explanation of the fact that
A is redder than B; see however \citet{Serra99} and \citet{Mich97}.
Next, the values of the flux ratio B/A differ in $R$ and $V$ (1.025
and 1.062, respectively).  This might be an effect of inadequate
galaxy modelling in the $V$-band (Sect.~\ref{S:qso_phot}), but it
seems very \emph{ad hoc} to us to re-adjust the galaxy photometry for
this reason only.  For comparison in the $R$-band, we note that in the
period 1992 October -- 1995 June (true dates corresponding to the
unshifted A image), the magnification ratio B/A was fairly stable at
$\sim$1.05, based upon the $R$-band data set treated by
\citet{OTSS03}.  The most straightforward explanation for this change
is that one or both quasar components are affected by (long-term)
microlensing.

Assuming that the observed gradient of, say, 3 mmag/day is typical for
the unsystematic brightness variations in QSO 0957+561, whatever their
origin, it follows from a simple random-walk argument that its
expected variability over a few weeks will be of the order of 0.02
mag, in good agreement with the investigation by \citet{Ullan03}. Over
a decade, the variations will be of the order of 0.2 mag, and such
amplitudes are easily found in the brightness record (e.g.\ 
\citealp{Pelt98,OTSS03}).

In conclusion, we hope that future observations and careful
photometric analysis of the data from the elusive QSO 0957+561 system
will shed light on some of the unresolved questions discussed above,
not to mention the issues of time delay, lens modelling, and source
structure.

\section*{Acknowledgments}
The work reported herein is based on observations made with the Nordic
Optical Telescope, operated on the island of La Palma jointly by
Denmark, Finland, Iceland, Norway, and Sweden, in the Spanish
Observatorio del Roque de los Muchachos of the Instituto de
Astrofisica de Canarias.  We thank L.~Goicoechea for exchanging ideas
with us regarding variability and photometric problems relating to QSO
0957+561. The observations were made at the NOT by K.~Kjernsmo,
\O.~Saanum and A.~K.~D.\ Evans.

\bibliographystyle{mn2e}
\bibliography{biblio}

\bsp
\label{lastpage}
\end{document}